\documentclass[a4paper, 10pt]{article}
\pdfoutput=1
\usepackage[cmex10]{amsmath}
\usepackage{amsfonts}
\usepackage{hyperref}
\usepackage{appendix}

\begin{document}
\title{A replay-attack resistant message authentication scheme using time-based keying hash functions and unique message identifiers}
\author{
    Boudhayan Gupta\\
    Department of Computer Science and Engineering\\
    NIIT University, Neemrana, RJ 301705, India\\
    Email: boudhayan.gupta@st.niituniversity.in
}
\maketitle{}


\begin{abstract}
    Hash-based message authentication codes are an extremely simple yet hugely effective construction for producing keyed message digests using shared secrets. HMACs have seen widespread use as ad-hoc digital signatures in many Internet applications. While messages signed with an HMAC are secure against sender impersonation and tampering in transit, if used alone they are susceptible to replay attacks. We propose a construction that extends HMACs to produce a keyed message digest that has a finite validity period. We then propose a message signature scheme that uses this time-dependent MAC along with an unique message identifier to calculate a set of authentication factors using which a recipient can readily detect and ignore replayed messages, thus providing perfect resistance against replay attacks. We further analyse time-based message authentication codes and show that they provide stronger security guarantees than plain HMACs, even when used independently of the aforementioned replay attack resistant message signature scheme.
\end{abstract}



\section{Introduction}
A very basic need in the domain of information security is to verify that sender of the message is not an impostor, and that the message was not tampered with while in transit. A common solution to this problem is to establish an encrypted communication channel between the sender and the recipient, using either symmetric key or asymmetric key cryptography. If the keys are known only to the sender and the recipient, it can be reasonably assumed that only the two of them can access messages being transmitted on the channel.

However, if an encrypted communication channel is not available, it is still possible to verify the authenticity and integrity of the sender and the message respectively, using a construct known as a \emph{Message Authentication Code (MAC)}. A MAC is generated from a message along with a cryptographic key, and is sent along with the message. Authenticating the message involves the recipient independently calculating the MAC from the message and the shared secret key and comparing it to that which was transmitted with the message. If the two codes match, the message is authentic. If they don't, then the message's claim to authenticity is disproven. Because MACs are keyed, an adversary cannot produce a new MAC for a tampered message without knowledge of the key.

While existing message authentication codes are resistant to both sender impersonation and tampering, they are not resistant to replay attacks. Even though the message may not be modified without failing authentication, the same message along with its MAC, if recorded by an adversary and re-sent to the recipient without any modification, will always authenticate successfully. A replay attack, while not as useful as tampering with the actual message, can have devastating consequences in certain circumstances, such as if the message being transmitted is some sort of command to perform a potentially destructive action.

A common method of guarding against replay attacks is to use an encrypted communication channel, thus denying the adversary access to the message in the first place. However, an encrypted communication channel makes MACs redundant, and fails to address the fundamental issue --- that of guarding against replay attacks in the presence of an adversary.

In this paper, we present a new construction, TMAC (Time-based MAC), as a special case of HMAC (Hash-based MAC). A TMAC is valid only for a specific time period, thereby reducing the amount of time during which replay attacks may be mounted to a finite period. We then present a new message signature scheme that uses a TMAC and an unique message identifier to calculate a set of authentication factors by which a recipient can specifically detect a replay attack and ignore the replayed message.


\section{Our Approach}
To make a message authentication code time-dependent, we can do one of two things. We can either make the message time-dependent, or we can make the validity of the key time-dependent.

Making the message time-dependent makes sense. In fact, a not-valid-before (NVB) and not-valid-after (NVA) date (or equivalent data) is present in many real-life ``messages'', including airline and train tickets, event invitations, and even digital signatures. As long as the NVB and NVA dates are also included in the portion of the message for which the MAC is generated, this is a perfectly valid way of creating time-dependent MACs.

Making the validity of the key time-dependent, however, is not only equally effective at creating time-dependent MACs, but also has critical advantages, foremost of which is that changing the key used to generate a MAC at regular intervals presents additional difficulties to attacks that aim to obtain the key. Simply obtaining the key isn't enough, as the key must be used during its validity period, and once it expires the new key has to be found once again. Making the key time-dependent forces a key-rotation schedule.

Of course, being forced to manually generate a new key after a fixed time interval may not always be desirable, as the new key must be shared with all parties involved in communication before it can be used. Therefore, we attempt to automate the key generation by using a pre-shared secret and a time-dependent key-derivation function to derive a time-dependent key from the master shared secret, an operation which can be done independently by all parties involved in communication.


\section{Preliminaries}
TMAC is defined as a special case of HMAC (Hash-based Message Authentication Code), and the key-derivation function we use is a simplified version of the Time-based One Time Password (TOTP) scheme. We shall revisit these schemes in the following two subsections.

\subsection{Hash-based Message Authentication Codes}
In~\cite{BCK1}, Bellare, Canetti and Krawczyk present a construction to produce keyed message authentication codes from public unkeyed cryptographic hash functions, by simulating the keying of the hash function's initialization vector. This construction is called HMAC, for Hash-based Message Authentication Code.

Given a cryptographic hash function \(H\), a cryptographic key \(K\), and a message \(m\), the HMAC function is defined as in Equation~\ref{equ_hmac_a}:

\begin{equation}
\label{equ_hmac_a}
    HMAC(K, m) = H((K \oplus opad) \| H((K \oplus ipad) \| m))
\end{equation}

where \({\oplus}\) denotes the bitwise exclusive-or operation, \({\|}\) denotes the concatenation operation, and \(opad\) (outer padding) and \(ipad\) (inner padding) are constants. The inner padding is the byte \(\mathtt{0x36}\) repeated enough times to make its length equal to the block size of the hash function \(H\). Likewise, the outer padding is the byte \(\mathtt{0x5c}\) repeated enough times to make its length equal to the block size of the hash function \(H\).

The HMAC construction is a rather good way of generating message authentication codes. Firstly, the construction ensures that only two extra rounds of the cryptohgraphic hash function are required to compute a MAC than to compute a plain hash, which for large messages is a negligible performance penalty. Secondly, this construction can be used with almost any cryptographic hash functions as the construction treats the hash function as a black box (In~\cite{BCK1} and~\cite{MB1}, proofs of security exist for when HMAC uses an underlying hash function that uses Merkle-Damg{\aa}rd construction. Proofs for hashes that use other constructions --- such as Sponge construction for Keccak and {SHA-3} --- do not exist in general; however the authors of Keccak make claims\cite{KECCAK} that HMACs using their hash function are as secure as HMACs using Merkle-Damg{\aa}rd hashes). Thirdly, as cryptographic hash functions are one-way functions, it is infeasible to derive the message, and therefore the key, from the {MAC}. Finally, it is infeasible to find a pair of messages that produce the same message authentication code, by virtue of the same property of cryptographic hash functions.

To authenticate a message using HMAC, the sender computes the HMAC for a message and sends it along with the message. The recipient independently calculates the HMAC for the same message and sees if the two HMACs match. If they do, the message is authentic. If they don't, the message has either been tampered with while in transit, or the sender is an impostor, or both. The key for the HMAC is never sent over the channel; it is known to both the sender and the recipient but not the adversary, and is established out-of-channel.

\subsection{Time-based One Time Passwords}
The Time-based One Time Password algorithm, as proposed by M'Raihi \emph{et.\ al.} in~\cite{RFC6238}, is a type of key-derivation function that takes a single secret (such as a password or passphrase) as an input and produces a single key as an output. This key is dependent on the secret as well as on the current time.

As defined in~\cite{RFC6238}, the algorithm is designed to produce a 6--8 digit human-readable decimal number. We do not need the output of our key-derivation function to be human-readable, and we certainly want a key that is longer than just 1--2 bytes (in their binary representation) or 6--8 bytes (when encoded as a string). Therefore, we can simplify the algorithm quite a bit by removing the truncation and modular addition steps.

The definition of the TOTP algorithm requires the use of HMACs. We begin by defining \(T_0\), a predetermined epoch relative to which all times are counted. Let \(T_s\) be a time step. Then we can derive a time counter \(T_C\) value like:

\begin{equation}
    T_C = \frac{Current\ Time - T_0}{T_s}
\end{equation}

The division by \(T_s\) is an integer division; we always throw away the remainder.

After we have derived the \(T_C\), we simply need to sign this with a HMAC to produce our TOTP key. Therefore, assuming \(K\) is our master secret, the complete definition of the TOTP function is thus:

\begin{equation}
    TOTP(K) = HMAC(K, T_C)
\end{equation}

The validity of the TOTP is at least 0 time units. Notice that because \(T_C\) is obtained by integer division, it does not change during a single validity interval. Therefore, the validity of the TOTP is at most \(T_s\) time units.


\section{Time-based Message Authentication Codes}
Now that we have revisited and defined the HMAC and TOTP functions, it is now possible to define the TMAC (Time-based Message Authentication Code) family of functions --- a family of functions that will produce a message authentication code that depends on both a secret key as well as the current time as authentication factors.

Let \(K\) be a shared secret used to generate the message authentication code, and \(m\) be the message to be signed. The generic TMAC function can be defined simply as:

\begin{equation}
    TMAC(K, m) = HMAC(TOTP(K), m)
\end{equation}

A complete definition of a particular TMAC function depends on the following parameters: \(H\), the cryptographic hash function used to generate the HMAC and therefore {TMAC}; \(T_0\), the epoch relative to which all times are counted; and \(T_s\), the time step for the underlying TOTP function. As an example, a particular instance of the TMAC function can be named \texttt{TMAC-SHA256-UNIX-30}, which denotes a TMAC generated using the SHA2-256 hash function, using the Unix epoch as \(T_0\) with a time step \(T_s\) of 30 seconds.

We suggest the use of the Unix epoch and a time step of 30 seconds as default for the function. Particular functions using these defaults may omit the epoch and time-step details from their function names and simply be named \texttt{TMAC-<hash name>}, e.g., \texttt{TMAC-SHA256}.

The validity of the TMAC is linked to the validity of the underlying {TOTP}. The time step for the TOTP is the time step for the TMAC, and the TMAC is valid for as long as the TOTP is valid. As in HMACs, we treat the actual cryptographic hash function underlying TMAC function as a black box; therefore the TMAC function can be defined using any cryptographic hash function.


\section{Unique Message Identifiers and Replay-attack Prevention}
While a TMAC allows control over the amount of time during which the MAC is valid, it alone is unable to prevent a replay attack occurring during the MAC's validity period. Therefore, an additional authentication factor is required to prevent replay attacks. In this section, we describe such a factor and define a scheme using said factor.

\subsection{Axioms}
We state that \textbf{signing the cryptographic hash of a message is equivalent to signing the message itself}. This statement is true for real-world hash functions because of the difficulty of finding a collision. For an unique message, the cryptographic hash can be considered to be unique.

\subsection{The scheme}
Let \(K\) be the secret key used to sign the message, and let \(m\) be the message we want to sign. We first generate a random string \(s\), which we use as an unique message identifier. The final signature is calculated in two steps:

\begin{enumerate}
    \item Let \(I = HMAC(s, m)\), an intermediate HMAC signature produced by signing the message with the unique message identifier as a key.
    \item Let \(S = TMAC(K, I)\), the final TMAC signature produced by signing the intermediate signature with the key.
\end{enumerate}

Let there be \(n\) messages, denoted as \(m_1, m_2, m_3 \ldots m_n\), their respective unique message identifiers be denoted as \(s_1, s_2, s_3 \ldots s_n\), and their respective message authentication codes be denoted as \(a_1, a_2, a_3 \ldots a_n\). To clarify, for any \(i \in [1, n] \cap \mathbb{N}\), the message authentication code for \(m_i\) is \(a_i\), and the unique message identifier is \(s_i\). To send a message authenticated by a TMAC, the sender transmits a 3-tuple \((m_i, a_i, s_i)\) using the channel. Therefore, for \(n\) messages, the channel carries, and the adversary sees, \((m_1, a_1, s_1), (m_2, a_2, s_2), (m_3, a_3, s_3) \ldots (m_n, a_n, s_n)\).

The recipient recieves the 3-tuple \((m_i, a_i, s_i)\), authenticates the message, and then \emph{retains \(s_i\)}. If the adversary tries to mount a replay attack, the recipient needs to refer to their list of already received \(s_i\) to detect whether a replay attack is in progress. If it is, the recipient discards the message.

Using an unique identifier for every message in itself is enough to prevent replay attacks, so why do we need to use a {TMAC}? The answer lies in the fact that without a TMAC, for true replay-attack resistence the recipient would need to retain the identifier to every single message ever recieved infinitely. Notice that the TMAC signs both the message and the identifier --- as soon as the TMAC becomes invalid, so does the identifier. This means that as soon as the time counter increments, the entire list of unique identifiers recieved by the recipient can be flushed, and the identifiers can be reused.

This means that the scheme can be modified to use, instead of randomly generated strings as message identifiers, a set of message identifiers assigned to the sender by the recipient. This can even be used to rate-limit the sender, as one identifier can only be used once during one time counter interval.

\subsection{Security analysis}
If an adversary tries to mount a replay attack, they have enough information to compute \(I\) for a message. The message is available, and the adversary can compute a new random string and compute the required {HMAC}. However, they can't compute the final TMAC, because to do that they would need the secret key.

Now what happens if the message is replayed once the time counter is incremented? The recipient has flushed their list of recieved message identifiers, therefore the replay attack is not specifically detected and the recipient attempts to authenticate the message. However, because the time counter has been incremented, the second step will generate a different TMAC signature for the recipient than that transmitted by the sender. The message will hence fail validation.

Because a TMAC cannot be forged without the key, the adversary cannot simply change the message identifier and compute a new {TMAC}.

\subsection{Implementation considerations}
An incorrect implementation of this scheme can result in the availability of a vector for a denial-of-service attack.

There is a single implementation rule that must be followed: the recipient must retain the unique message identifier for the message \emph{only after authenticating the message}.

If the sender wishes to send the recipient a message, the adversary can transmit a dummy message with invalid message authentication codes but the same message identifier as the legitimate message to the recipient. If the adversary's message arrives at the recipient before the legitimate message, and the recipient does not wait until the message is authenticated before retaining the unique message identifier, then when the legitimate message arrives, it will be falsely classified as a replay attack and rejected.

Using this implementation rule also makes implementing parallel recipients easier. A na\"{\i}vely implemented parallel authenticator might introduce a race condition whereby the identifier for a fabricated message is retained before the identifier for a legitimate message can be retained. Waiting for authentication before retaining the identifier will alleviate this race condition. Remember that there can only be one legitimate message for a message identifier.

It is therefore imperative that the recipient wait until the message is authenticated before retaining the unique message identifiers.


\section{Security Analysis of TMAC}
In this section, we present a security analysis of Time-based Message Authentication Codes. We prove that the best possible attack against this scheme is the birthday attack, and also show that a practical birthday attack on TMAC is extremely difficult.

\subsection{Assumptions}
We perform this security analysis under the assumption that the adversary does not know \emph{a priori} that the authentication scheme in use is {TMAC}, but can recognize the use of {HMAC} in the message authentication scheme. The absence of this assumption does not have a negative implication on the real security of the scheme, because recognizing the use of {TMAC} is as trivial as performing chosen-plaintext attacks with the same plaintext for a time duration during which the time counter is incremented.

\subsection{Birthday attacks}
Let the messages be denoted as \(m_1, m_2, m_3 \ldots m_n\), and the respective message authentication codes be denoted as \(a_1, a_2, a_3 \ldots a_n\), where for any \(i \in [1, n] \cap \mathbb{N}\), the message authentication code for \(m_i\) is \(a_i\). To send a message authenticated by a TMAC, the sender transmits a 2-tuple \((m_i, a_i)\) using the channel. Therefore, for \(n\) messages, the channel carries, and the adversary sees, \((m_1, a_1), (m_2, a_2), (m_3, a_3) \ldots (m_n, a_n)\).

Without any \emph{a priori} knowledge of the authentication scheme in use, the adversary's ultimate goal is to try and discover a function \(B(x)\), such that:

\[
    a_i = B(m_i)
\]

If the adversary manages to obtain such a function, they can sign any message and produce a valid message signature. Thankfully, this is an extremely difficult attack to perform; an easier attack to is to find another message \(m_j\) that also produces the same message digest \(a_i\). If such a message is found, it can be substituted in place of the original message and still pass validation.

Recall that the TMAC is simply a special case of {HMAC}, and that an HMAC is simply a cryptographic hash function whose initialization vector is a secret key. Therefore, any attack that finds a collision on the underlying hash function \emph{should} find a collision on the HMAC, and by extension {TMAC}.

Let the message authentication code be \(c\) bits in size. Therefore, by the pigeonhole principle, a collision (two messages producing the same MAC) is guaranteed to be found if the MAC for \(2^c + 1\) messages are computed. However, the solution to the birthday problem greatly simplifies the problem. In general, the probability that a collision will be found reaches \(50\% \) with around \(1.2 \times \sqrt{(2^c)}\) trials, which we simplify by saying that we expect to find a collision after calculating the hashes for \(2^{\frac{c}{2}}\) messages.

Note that finding a collision does not imply that the entire authentication scheme is broken. In particular, the second message that produces the same message digest may be meaningless to the recipient, and be rejected on grounds of being gibberish.

We now prove that theoretically, this is the best attack possible on a TMAC, and that practically, a birthday attack on a TMAC is extremely difficult to perform.

\subsection{Brute force are the best attacks}
The birthday attack places a bound on the complexity of the best possible brute-force collision-finding attack on any hashing scheme. If we can find a better attack (i.e., an attack that has a greater probability of finding collisions in fewer trials), we consider the scheme to be broken.

In~\cite{BCK1}, it is proved that a break in the HMAC construction constitutes a break in the underlying hash function. In~\cite{MB1}, it is further proved that HMAC is secure even if the underlying hash function is only weakly collision resistant. There are no known attacks that break the HMAC scheme itself. Therefore, the assumption that HMAC is secure currently holds true.

In~\cite{RFC4226}, a proof to show that brute-force attacks are the best attacks on the HOTP scheme exists. TOTP is a specific case of HOTP, and therefore the same proof applies to show that brute-force attacks are the best attacks on the TOTP scheme.

Recall that TMAC is defined as:

\begin{equation}
\label{equ_tmac_a}
    TMAC(K, m) = HMAC(TOTP(K), m)
\end{equation}

And that TOTP is defined as:

\begin{equation}
\label{equ_totp_a}
    TOTP(K) = HMAC(K, TC)
\end{equation}

Therefore, from equations~\ref{equ_tmac_a} and~\ref{equ_totp_a}, we can re-write the equation for TMAC as:

\begin{equation}
\label{equ_tmac_b}
    TMAC(K, m) = HMAC(HMAC(K, TC), m)
\end{equation}

This essentially shows that we are using a value randomly picked from an uniform distribution as the key for the outermost application for the {HMAC}. As long as the HMAC scheme itself isn't broken, TMAC remains secure.

\subsection{Time requirements for birthday attacks}
To prove that a practical birthday on TMAC is difficult to perform, we'll need to take a closer look at the mathematics behind the birthday attacks on a hash function.

An \(n\)-bit cryptographic hash function is a function that takes a message of any length and produces a \emph{hash} that is exactly \(n\) bits in length. Formally, a hash function \(H\) is defined over a binary alphabet and maps any binary string to an \(n\)-bit binary string:

\begin{equation}
    H_n: {\{}0,1{\}}^* \to {\{}0,1{\}}^n
\end{equation}

The set of all possible output values of the hash function is the set of all possible \(n\)-bit binary strings. The number of possible \(n\)-bit binary strings is \(2^n\). An ideal hash function outputs completely random hash values, which means the hash function's output is uniformly distributed across all possible output values (i.e., all possible output values are equally likely). Real-world hash functions are well balanced, i.e., their output is close to uniformly distributed.

The birthday problem states that if there are \(m\) possible hash values and we hash \(k\) messages to produce their hashes, the probability that a collision will occur, i.e., the probability that we will find a pair of messages \(x_i\) and \(x_j\) such that \(x_i \neq x_j\) and \(H_n(x_i) = H_n(x_j)\) is given by the expression:

\begin{equation}
\label{equ_hash_m}
    P(m, k) = 1 - \frac{m!}{(m - k)!m^k}
\end{equation}

For a \(c\)-bit binary hash value, we can re-write this function as:

\begin{equation}
\label{equ_hash_c}
    P(c, k) = 1 - \frac{2^c!}{2^{ck}(2^c - k)!}
\end{equation}

Now let us assume that we have access to a machine that can compute the hashes for \(d\) messages per second. In \(t\) seconds, it can calculate the hashes for \(dt\) messages. Therefore, we can redefine Equation~\ref{equ_hash_c} as:

\begin{equation}
    P_{c,d}(t) = P(c, d, t) = 1 - \frac{2^c!}{2^{cdt}(2^c - dt)!}
\end{equation}

We're interested in a value of \(t\) in terms of \(d\) such that for a \(c\) bit hash value, the value of \(P_{c,d}(t)\) reaches \(0.5\). This is the amount of time that our machine needs to run in order to perform a successful birthday attack.

This is where we start approximating values to make our work easier. We can use the solution to Taylor's expansion series of \(e\) to approximate Equation~\ref{equ_hash_m} as:

\begin{equation}
\label{equ_hash_appx}
    P(m, k) \approx 1 - e^{\frac{-k(k-1)}{2m}} \approx 1 - e^{\frac{-k^2}{2m}}
\end{equation}

Let \(k(P, m)\) be the smallest number of \(k\) that we have to choose so that the probability of finding a collision becomes at least \(P\). By inverting Equation~\ref{equ_hash_appx}, we find the following equation:

\begin{equation*}
    k(P, m) \approx \sqrt{2m\ln\left({\frac{1}{1-P}}\right)}
\end{equation*}

By assigning a value of \(0.5\) to \(P\), we arrive at this equation:

\begin{equation*}
    k(0.5, m) \approx 1.1774\sqrt{m}
\end{equation*}

In the case of a \(c\) bit hash function running on a machine which is hashing \(d\) messages per second and running for \(t\) seconds, the expression becomes:

\begin{equation*}
    k(0.5, c) = dt \approx 1.1774 \times 2^{\frac{c}{2}}
\end{equation*}

We can re-arrange this to obtain:

\begin{equation}
\label{equ_hash_time}
    T_c(d) = t = \frac{1.1774 \times 2^{\frac{c}{2}}}{d}
\end{equation}

The function \(T_c(d)\) gives us the amount of time in terms of the hashrate \(d\) for which the machine has to run to find a collision for a \(c\)-bit hash function using a birthday attack.

In the next section, we will use this function to compute the time required to run a practical birthday attack and show that it is far in excess of practical values of the time step used to compute a {TMAC}.

\subsection{Birthday attacks may not always be feasible}
The validity of the proof that a birthday attack on a cryptographic hash function increases the probability of finding a collision assumes that the hash function will always produce the same message digest for a given message. However, while TMAC is essentially also a cryptographic hash function with a secret initialization vector, the said initialization vector is not constant, and is a function of the current time. Therefore, at different points in time, the TMAC for the same message will be different.

Let us assume that the adversary does not know that the authentication scheme in use is TMAC, and proceeds to perform chosen-plaintext attacks to gather a corpus of \((m, a)\) pairs with which to find a collision. We know that the birthday problem greatly reduces the number of trials required to find a collision.

Let us also assume that the time step for the TOTP function is sufficiently large so that the required number of \((m, a)\) pairs can be collected to perform a birthday attack without the underlying hash function changing. This essentially reduces the security of TMAC to that of {HMAC}. How large does this time step have to be?

We defined a function \(T_c(d)\) in Equation~\ref{equ_hash_time} to find the answer to exactly this problem. But before we can compute a value of the time required, we must decide what practical values for the hashrate \(d\) and the hash length \(c\) are.

The absolute worst hash function that is still regarded by the cryptographic community as being ``good enough'' to compute HMACs is SHA-1. SHA-1 hashes are 160 bits wide, and therefore we expect to find a collision by hashing \(2^{80}\) messages. This expectation is not too far from reality, because SHA-1 has been shown to be highly balanced\cite{MT1} and thus has a close to ideal collision resistance. Therefore, we will set the value of \(c\) at 160.

To find a realistic value of \(d\) is a little more difficult. It has been said that large intelligence agencies supported by well-funded governments already have machines capable of finding collisions in SHA-1 in a reasonable amount of time\cite{EET1}. However, even if we restrict ourselves to commercial hardware, finding a realistic value of \(d\) is still difficult, because even on an average machine, the time required to hash a message still depends on the length of the message.

A good place to look for the state of the art in hashing hardware is Bitcoin mining. Bitcoin hashes are double {SHA2-256}, and application-specific integrated circuits (ASICs) exist that can compute hashes extremely fast. The fastest such hardware, the AntMiner S5+\cite{AMS5}, can compute 7722 billion hashes per second. We will use this number for \(d\), as it represents the absolute worst case for time bounds in finding hash collisions.

With \(c = 160\) and \(d = 7722 \times 10^9\), the value of \(t_s\) comes out to be:

\begin{align*}
    t_s &= T_{160}(7722 \times 10^9)\\
        &= \frac{1.1774 \times 2^{80}}{7722 \times 10^9}\enskip seconds\\
        &= 184.33 \times 10^9\enskip seconds\\
        &= 5841.16\enskip years
\end{align*}

What can we infer from this?

Because for the purposes of performing a birthday attack on the TMAC we're treating it as a plain hash function, we're relying on the assumption that the hash function will always produce the same hash for the same message. As long as the time counter doesn't change, this assumption holds true. As soon as the time counter increments, however, the message that is being hashed by the outermost application of the hash function (refer to Equation~\ref{equ_hmac_a}) changes. This essentially means that if the time counter changes, the hash function for which we're trying to find a collision ceases to exist, making the entire attack meaningless.

Realistic values of the time step are expected to be orders of magnitude lesser than nearly six thousand years. For example, in~\cite{RFC6238}, which describes the use of TOTPs in two-factor authentication, the default and suggested time step is 30 seconds. Advances in computer hardware and software are expected to increase hashrates on average computers exponentally in the coming years; this might need to be mitigated by either choosing longer hash functions or shorter time steps.


\section{Conclusion}
In this paper, we have described a simple and secure method for computing time-dependent message authentication codes which along with unique message identifiers can be used to prevent replay attacks on messages passed through an insecure channel. The method is directly practically applicable and does not require unrealistic hardware, such as an infinite amount of memory. It is also much simpler, as well as less computationally expensive than full end-to-end channel encryption.

The security analysis proves that TMAC is at least as secure as HMAC, and that in all real-world scenarios TMAC is actually harder to crack. Since both simple HMACs and TMACs require the same input arguments (a secret key and a message), we recommend that stakeholders currently using HMACs to sign messages consider using simple TMACs instead, without the unique message identifier based replay-attack prevention mechanism, as a drop-in replacement for HMACs.


\begin{appendices}
\section*{Acknowledgment}
The author would like to thank Dr.\ Debasis Das for his invaluable help in mentoring him during the process of writing this paper. The author is also thankful to Dr.\ Kumar Vishal for his help with understanding the mathematics behind birthday attacks.
\end{appendices}



\end{document}